# SAIL: Machine Learning Guided Structural Analysis Attack on Hardware Obfuscation

Prabuddha Chakraborty, Jonathan Cruz, and Swarup Bhunia
Department of Electrical & Computer Engineering
University of Florida, Gainesville, FL, USA

*Abstract*— Obfuscation is a technique for protecting hardware intellectual property (IP) blocks against reverse engineering, piracy, and malicious modifications. Current obfuscation efforts mainly focus on functional locking of a design to prevent black-box usage. They do not directly address hiding design intent through structural transformations, which is an important objective of obfuscation. We note that current obfuscation techniques incorporate only: (1) local, and (2) predictable changes in circuit topology. In this paper, we present SAIL, a structural attack on obfuscation using machine learning (ML) models that exposes a critical vulnerability of these methods. Through this attack, we demonstrate that the gate-level structure of an obfuscated design can be retrieved in most parts through a systematic set of steps. The proposed attack is applicable to all forms of logic obfuscation, and significantly more powerful than existing attacks, e.g., SAT-based attacks, since it does not require the availability of golden functional responses (e.g. an unlocked IC). Evaluation on benchmark circuits show that we can recover an average of around 84% (up to 95%) transformations introduced by obfuscation. We also show that this attack is scalable, flexible, and versatile.

## I. Introduction

Hardware intellectual property (IP) based system-on-chip (SoC) design has become a prevalent practice in the semiconductor industry. However, the global economic trend that dictates a horizontal business model incorporates many untrusted parties in the modern chip design flow. In particular, most chip designers rely on untrusted 3-rd party fabrication facilities. Such a trend diminishes a chip designer's control on the IPs and makes them vulnerable to various forms of attacks, including piracy, reverse-engineering, overproduction, malicious modifications, or Trojan attacks, leading to serious economic and security threats [11] [3]. To address these security issues, hardware obfuscation techniques have been actively studied for the past decade. They aim at transforming a design - both functionally and structurally based on a key, such that an obfuscated design functions correctly only if the right key inputs are provided. Fig. 1, summarizes the two main goals for obfuscation: 1) preventing black-box usage, and 2) hiding design intent. Using functional locking techniques such as XOR gate insertion locking scheme [5], [6], [7], black-box usage of the IP/IC can be prevented. However, an attacker may also be interested in understanding the intent of a design through reverse engineering of the netlist. In order to address this important need, judicious structural transformations need to be introduced in a design. Current obfuscation approaches primarily rely on: 1) insertion of various modification cells (e.g., XOR, XNOR gates) controlled by key bits, and 2) re-synthesis of the design after obfuscation, to obtain any structural changes in the design.

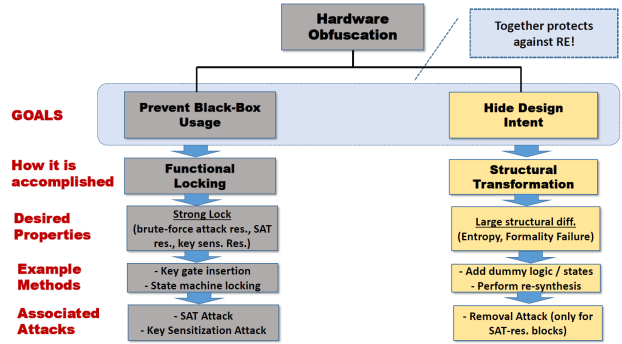

Fig. 1: Primary objectives of Hardware Obfuscation.

Like any security solutions, IP protection achieved through obfuscation largely depends on its robustness against possible attacks. Over the years, several attacks on obfuscation have been proposed, which broadly fall into two categories: 1) *functional attacks*, such as SAT attack [2], and 2) *structural attacks*, such as ANTI-SAT block removal attack [4]. However, most reported attacks on obfuscation have been functional attacks. The only structural attack [4], is actually not designed to deobfuscate an IP, but to facilitate a subsequent SAT attack [2]. Based on a thorough statistical analysis, we make two key observations regarding structural changes: 1) obfuscation introduces sparse and local structural changes in a design, and 2) the changes are very deterministic (i.e., follow a set of well-known logic synthesis rules). Motivated by these observations, we introduce a new paradigm of attack on obfuscation, called **SAIL** (Structural Analysis using Machine Learning), which exposes a critical vulnerability in logic obfuscation approaches. It can retrieve the original design and hence, the design intent, through structural analysis guided by machine learning. Given an obfuscated design, we extract each obfuscation gate locality (subgraphs, considering netlist as a graph) and revert them to their pre-synthesis state using the reconstruction neural network trained on 11160 [Pre-Synthesis, Post-Synthesis] locality pairs. Using this information gained on the pre-synthesis obfuscated netlist, an attacker can more easily carry out key guessing and reverse engineering attacks by observing the structure. We also use a change prediction ML model to improve accuracy and reduce computation time.

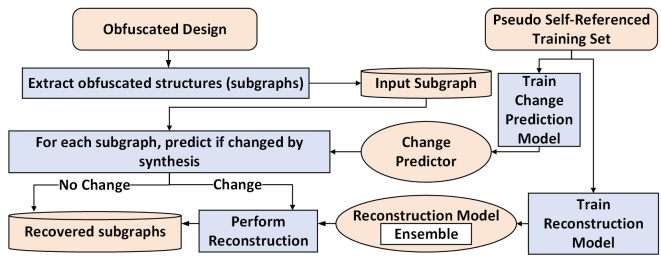

Fig. 2: Overview of major steps in the SAIL Attack Flow.

SAIL is a more powerful attack than SAT [2] based functional attacks due to the following reasons: (1) unlike SAT, SAIL does not require golden responses, e.g., from an unlocked IP, which may be difficult to obtain for an attacker, 2) it can be applied to both combinational and sequential designs, 3) unlike SAT attack, which fails for specific functions, e.g., multiplier, it works well for all designs and does not depend on the underlying boolean function, and finally, 4) it is scalable in terms of accuracy and computation time with respect to both key and benchmark size. We quantitatively and qualitatively analyze the effectiveness of SAIL attack using the systematic framework and tool that we have developed as shown in Fig. 2. From our results, we demonstrate that SAIL attack is very effective against different key gate insertion heuristics and key sizes. We achieve an average of 84.14%, and up to 94.98% recovery of the obfuscated structures for a set of benchmark circuits.

In particular, we make the following key contributions:
- We analyze, both quantitatively and qualitatively, the nature of structural changes in a gate-level netlist introduced due to different steps of obfuscation. We present the salient observations in this regard that create the foundation of the proposed attack.
- We present a complete framework of SAIL attack with systematic set of steps that builds on this analysis. It includes (1) *Change Prediction Model* that can predict whether the obfuscation gates have undergone changes due to obfuscation; and (2) a *Reconstruction Model* that can locally revert the structural changes post re-synthesis.
- We also present construction of a *Change Prediction Boosted Reconstruction Model* using both the *Reconstruction Model* and the *Change Prediction Model* for improved recovery.
- Finally, we present comprehensive evaluation of the attack on benchmark circuits and demonstrate that it is scalable with respect to the key length, different key-gate insertion heuristics, and design size.

## II. Background and Related Works

Logic obfuscation attempts to hide the functionality of a design by inserting gates that act as a lock and are controlled by key inputs. Functional behavior in the design is only restored upon entering the correct key. In general, there are two types of logic locking: combinational and sequential. For combinational logic locking, combinational gates (XOR/XNOR, AND, etc.) are inserted into a design with one input serving as a key-bit. To lock a design with an $N$-bit key at least $N$ combinational key-gates must be added. On the other hand, sequential logic locking modifies the finite state machine by adding additional key-state transitions. These transitions are properly traversed by applying the correct sequence of key-bits and subsequently unlock the design. The focus of this paper is on combinational logic locking. In this obfuscation scheme [5], [6], [7], a design is locked to hide functionality with an $N$-bit key by inserting appropriate key gates (e.g., XOR for key-bit '0' and XOR followed by an inverter for key-bit '1').

Most of the attacks on obfuscation try to retrieve the key, based on functional analysis of the design. A key sensitization attack tries to retrieve the key-bits by applying a pattern of non-key inputs such that a key-input is mapped to an observable output (sensitization) [13]. In [2], a SAT-based algorithm is proposed to completely or partially retrieve the key from a combinational design. It requires the obfuscated netlist and an unlocked IC/netlist is required to carry out the attack. Moreover, SAT formulation is an NP-Complete problem and the heuristic-based solutions are not guaranteed to give a solution within reasonable time. Different protections [9], [10], [8] against SAT attack [2] have been proposed that generally involve adding extra logic to increase either the number of iterations of the attack or the time to complete each iteration. In [1], an approximate key is retrieved that shows output corruptibility for very few inputs and can bypass the protections offered by [9]. To the best of our knowledge, the only structural attack on obfuscation [4] is aimed at removal of ANTI-SAT blocks [9] to facilitate a subsequent SAT attack. Conversely, our work aims at quantitatively and qualitatively analyzing and evaluating the effect of obfuscation on the structure of the gate-level netlist and the deterministic heuristics involved in structural hiding. The only ML based attack on obfuscation [12] is a key retrieval attack, which also requires an unlocked IC and is expected to suffer from scalability issues. In comparison, we propose a netlist structural analysis based attack that does not require golden responses and is very scalable.

## III. Motivation

In this section, we discuss two fundamental observations on obfuscation-induced structural changes and we justify the use of machine learning models for structural analysis based attacks.

### A. Obfuscation Induced Changes Are Local

When a netlist is obfuscated using a XOR-based locking scheme, a re-synthesis of the netlist is performed in an attempt to camouflage the key-gates that are inserted. In the best case, the key XOR-Gate itself will combine with local gates and transform into a new structure or cease to be connected to the Key Input. We refer to this as $Level-3$ change. In a slightly worse case, the XOR-Gate inserted due to obfuscation may remain intact but its neighboring gates may change due to logic simplification. We call this a $Level-2$ change. Finally, the third scenario observed involves no change of the inserted obfuscation gates or surrounding gates after re-synthesis, referred to as $Level-1$ changes. As a result, the obfuscated design remains vulnerable to removal and key guessing attacks. Moreover, we observe that this process does not aid in hiding the design intent and fails to obfuscate the design from a reverse engineering standpoint.

In Table I, for 3720 gate localities, we note 36.5% of key inserted localities do not go through any structural change. Another 57.31% of the localities only show Level-2 changes which are very easy to revert using certain rules which can be manually devised or learned through statistical methods. Only the remaining 6.18% of the localities are properly obfuscated. However, some of these changes are predictable and can be locally reverted as well with the help of statistically generated rules/algorithms.

TABLE I: Number of each type of changes for IPs.

|  | Level-1 | Level-2 | Level-3 |
|---|---|---|---|
| **c1355** | 33 | 87 | 0 |
| **c1908** | 31 | 88 | 1 |
| **c2670** | 45 | 66 | 9 |
| **c3540** | 205 | 256 | 19 |
| **c5315** | 321 | 562 | 77 |
| **c6288** | 432 | 482 | 46 |
| **c7552** | 291 | 591 | 78 |
| **Avg** | **1358 (36.50%)** | **2132 (57.31%)** | **230 (6.18%)** |

TABLE II: Limited number of rules govern most changes.

| # Rules | % Change | #Rules | % Change | #Rules | % Change |
|---|---|---|---|---|---|
| **1** | 10.69 | **38** | 70.29 | **180** | 90.02 |
| **6** | 41.07 | **81** | 80.00 | **290** | 95.00 |
| **11** | 51.02 | **120** | 85.00 | **476** | 100.00 |

In Fig. 3(a)-(d), we show the most common Level-2 and Level-3 changes that we have observed during our experiments and the resulting localities after we perform the reconstruction. In Fig. 3(a), we see the inverter moving from behind the XOR gate to its front. This transformation is a very common scenario and can confuse the reverse engineer into thinking the *Inverter* is not an obfuscation gate. By observing the locality (up to 10 gates) around the key-Gate, our statistically learned rules can determine the state of the locality before synthesis and recover it, thereby removing confusion. In Fig. 3(b), we observe a very common level-3 change, where the inverter comes between the key input wire and the XOR gate. This change can also be easily recovered and stitched back into the design. Fig. 3(c) and Fig. 3(d) depict more complex level-3 changes but our models are still able to locally recover from such changes.

### B. Why Machine Learning? The Synthesis Tool Optimization is Deterministic

In the previous section, we have statistically established that the changes induced by obfuscation gates are local and limited. We can go one step further and try to recover, from these limited changes, a local snapshot of the key-gate inserted region. Recovering a small locality around the obfuscation gate connected to the key input wire is enough to obtain insight into the obfuscation that was carried out.

Given the designs before and after synthesis, we enumerate each unique transformation (each unique [Input Locality, Output Locality] pair) carried out by the synthesis tool. We observe that very few rules are used by the synthesis tool for carrying out most of the transformations. In Table II, we observe 10.69% of the transformations are done with only one rule. With six rules, the synthesis tool does 41.07% of the transformations and only 180 rules govern 90.02% of the changes. If we can statistically learn these limited number of rules then we can revert the changes introduced by them. This is the main reason why a machine learning reversion attack is possible. Table II statistics is based on observing 3720 different localities (of size 3 gates) across multiple iterations of obfuscations over the 7 largest ISCAS-85 benchmarks.

The number of types of change is limited and that alone allows a learning process to generate the required rules for reversion given enough statistical data. This claim is further corroborated by the quantitative recovery results in Section V. In the next section, we will introduce our recovery attack models in details.

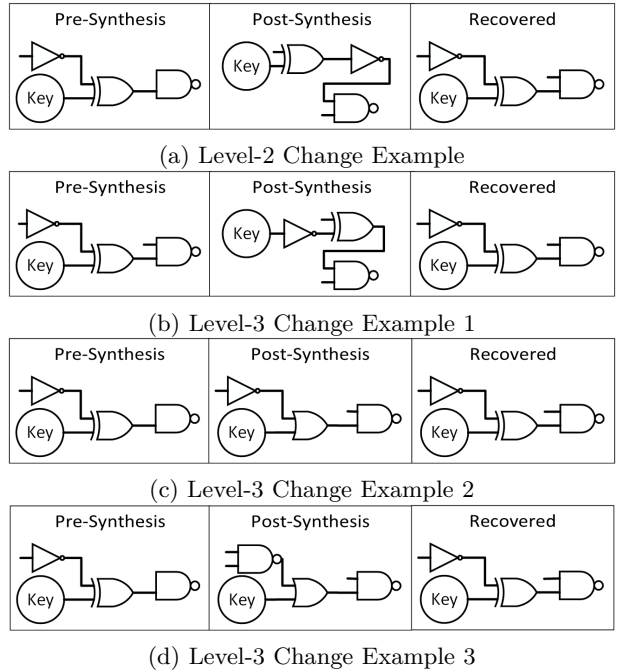

(a) Level-2 Change Example

(b) Level-3 Change Example 1

(c) Level-3 Change Example 2

(d) Level-3 Change Example 3

Fig. 3: Example recovery using our model.

## IV. Attack Methodology

In this section, we shall formulate the three attack models in subsections IV-B and IV-C.

### A. Training Dataset Generation

Through extensive analysis, we observe that our models work best if they are trained specifically for a particular circuit. As we do not have access to the original pre-obfuscation netlists, we take the obfuscated circuit provided to the attacker and treat it as a *Pseudo Golden Circuit* and carry out one more round of obfuscation to create the training set. This way, we can capture the circuit-specific information and at the same time eliminate the need for a *Golden Circuit*. It is important to note that the gates inserted due to the original obfuscation are a minor fraction of the total number of gates in the circuit and hence do not have much of an impact on the subsequent obfuscation. We term this method *Pseudo Self Referencing*.

### B. Change Prediction Model

As most of the obfuscation gate localities do not change due to the re-synthesis process, we propose a *Change Prediction Model* which can determine with good accuracy which localities change and which do not. The model is trained using a dataset containing [Locality, Change Indicator Boolean] pairs constructed using the pseudo-self referencing scheme. Once trained, the model can be used to predict whether or not a locality in the test set (original obfuscated designs) is changed due to synthesis. We use a Random Forest as the ML model for our experiments. We have tried SVM, Logistic Regression, and several other ML models, but Random Forest gives the best results in terms of both computation efficiency and accuracy.

*C. Reconstruction Model*

The Reconstruction/Reversion Model is used to locally revert the changes caused by the synthesis performed after the obfuscation. The model is trained using a dataset containing [Post-Synthesis Locality, Pre-Synthesis Locality] pairs constructed using the pseudo-self referencing scheme. Once trained, the model can be used to predict the Pre-Synthesis Locality given a Post-Synthesis Locality from the test set (original obfuscated designs). The model used for obtaining the results is a multichannel, multilayer neural network. To further improve the efficiency, we train multiple such models for varying Post-Synthesis Locality sizes (from 3 gates locality up to 10 gates locality) and combine them using a standard cumulative confidence voting *ensemble* scheme as seen in Fig. 2.

Although the Reconstruction/Reversion model works well on it own, the accuracy can be further boosted by using the change prediction model to determine which localities to reconstruct and which localities to leave alone.

## V. Results and Analysis

To evaluate the effectiveness of our attack we apply XOR logic locking on several ISCAS-85 benchmarks. We generate training and testing sets of 11160 samples and 3720 samples, respectively. To generate such a big training set, we obfuscate each benchmark multiple times in separate instances using random key-gate insertion heuristics using the tool provided in [11]. In each instance [8,8,8,32,64,64,64] bit keys are inserted for c1355, c1908, c2670, c3540, c5315, c6288, and c7552, respectively. The following sections describe our results in detail.

*A. Change Prediction Model*

By looking at the post-synthesized design, we can statistically determine if the key-gate inserted locality underwent any change. This is possible only because the synthesis tool is not random and exhibits deterministic nature. This effect is further exacerbated for designs that have regular repetitive structures such as *c6288* which shows 98.43% change prediction accuracy for input locality size of 10 gates. On average, all benchmarks show 81.76% accuracy for 10 gates input locality size showing us that the method works for a varied range of designs. Fig. 4, shows the accuracy of the model for different benchmarks and the effect of input locality size on the accuracy. The average is shown in *"Red"* clearly indicates that the accuracy increases with the input locality size and plateaus after locality of size 6 gates.

*B. Reconstruction Model*

To evaluate the reconstruction model, we analyze how much of the changes due to obfuscation can be recovered locally considering an output locality size of 3 gates, referred to as a *snapshot*. We can correctly recover many changes as evident from Table III. In case of 1 to 2 incorrect gate(s) in the snapshot (Gate Error = 1 and Gate Error = 2), the snapshot prediction is partially incorrect but remains useful for subsequent analysis. We propose a metric $R$ as shown in Eqn. 1. GE[x] stands for $GateError = x$. A snapshot with 1 gate type wrong has a weight of 2/3 and the snapshot with 2 gate types wrong has a weight of 1/3. Link Error is defined as the number of incorrect connections between the gates in the predicted locality. The ensemble we use as the final reconstruction model is made from combining several models, separately trained using datasets with different input locality (Post-Synthesis locality) sizes. As seen in Fig. 5, for different benchmarks the optimal Post-Synthesis locality size varies and the ensemble is the best way to stabilize the variance.

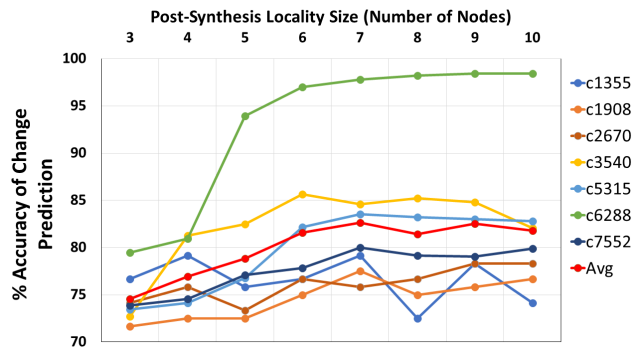

Fig. 4: Accuracy of change prediction model.

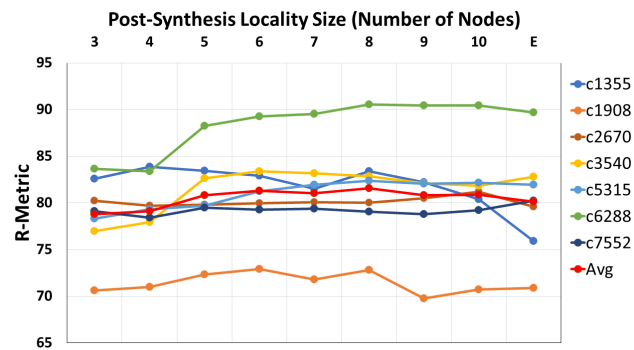

Fig. 5: Effect of Post-Synthesis locality size on the R-Metric for different benchmark circuits. E: Ensemble.

$$R = (GE[0] * 1) + (GE[1] * 0.66) + (GE[2] * 0.33) \quad (1)$$

In Table IV, we observe the amount of complete recovery (Gate Error = 0 and Link Error = 0) for each type of changes. For Level-1 changes (no change) 17.4 % of the localities are incorrectly modified. To reduce this error, we use the Change Prediction model. From Table V, we observe a 10.86% increase in the complete recovery accuracy (G = 0, L = 0) and around 4% increase in R-Metric. In Table VI, the complete recovery accuracy (G = 0, L = 0) for different types of changes are shown. Around 12% more of the Level-1 changes (i.e. no change cases) are correctly recovered. In certain cases, the locality structural changes are due to oscillating logic simplifications over several consecutive re-synthesis steps. However, these changes are not induced by the inserted key-gates as the key-gates are intact in their original form. Therefore, we do not need to do any reconstruction/reversion for such localities. The learned change prediction algorithm is able to detect those cases and mark them as *no change*. As a result, we see $\sim 9.6\%$ hike in the level-2 change (simple local change) accuracy.

TABLE III: Accuracy for Reconstruction Model without Change Prediction boost. G: Gate Error, L: Link Error.

|  | G=0 L=0 | G=1 L=0 | G=2 L=0 | R-Metric |
|---|---|---|---|---|
| **c1355** | 57.49 | 25.83 | 4.16 | 75.91 |
| **c1908** | 47.50 | 32.50 | 5.83 | 70.87 |
| **c2670** | 70.00 | 12.50 | 4.16 | 79.62 |
| **c3540** | 65.62 | 23.75 | 4.58 | 82.80 |
| **c5315** | 67.50 | 19.47 | 4.79 | 81.93 |
| **c6288** | 78.54 | 12.70 | 8.43 | 89.70 |
| **c7552** | 58.95 | 28.64 | 7.29 | 80.25 |
| **Avg** | **63.65** | **22.19** | **5.60** | **80.15** |

TABLE IV: Gate Error = 0, Link Error = 0 accuracy for Reconstruction Model (Without Change Prediction Boost) across different types of changes.

|       | Level-1          | Level-2          | Level-3        |
|-------|------------------|------------------|----------------|
| c1355 | 57.57 (19/33)    | 57.47 (50/87)    | NAN (0/0)      |
| c1908 | 77.41 (24/31)    | 37.50 (33/88)    | 0.00 (0/1)     |
| c2670 | 82.22 (37/45)    | 65.15 (43/66)    | 44.44 (4/9)    |
| c3540 | 80.48 (165/205)  | 55.85 (143/256)  | 36.84 (7/19)   |
| c5315 | 90.65 (291/321)  | 57.11 (321/562)  | 46.75 (36/77)  |
| c6288 | 86.34 (373/432)  | 72.40 (349/482)  | 69.56 (32/46)  |
| c7552 | 73.19 (213/291)  | 54.65 (323/591)  | 38.46 (30/78)  |
| Avg   | 82.6(1122/1358)  | 59.1(1262/2132)  | 47.3(109/230)  |

TABLE V: Accuracy for Change Prediction Boosted Reconstruction Model. G: Gate Error, L: Link Error.

|       | G=0 L=0 | G=1 L=0 | G=2 L=0 | R-Metric |
|-------|---------|---------|---------|----------|
| c1355 | 73.33   | 10.00   | 2.50    | 80.75    |
| c1908 | 66.66   | 14.16   | 6.66    | 78.20    |
| c2670 | 74.16   | 5.83    | 3.33    | 79.10    |
| c3540 | 73.95   | 17.08   | 4.79    | 86.80    |
| c5315 | 74.89   | 12.50   | 4.20    | 84.52    |
| c6288 | 87.50   | 10.62   | 1.45    | 94.98    |
| c7552 | 71.14   | 18.02   | 4.89    | 84.64    |
| Avg   | 74.51   | 12.60   | 3.97    | 84.14    |

In Fig. 6, the effectiveness of the model is visually represented. In Fig. 6 (a) the obfuscated design post re-synthesis may have gone through local optimizations around key-gates inserted localities so all such localities are marked as red nodes and cyan nodes represent the key input wires. Fig. 6 (b), shows the reconstructed design after recovery. Note, that using our model, we can only recover snapshots. The ability to automatically stitch snapshots back into the design depends on the type of transformation that the locality underwent. This process can be potentially difficult, especially when the pre and post snapshot input sizes differ. However, simply recovering a snapshot around the inserted key-gates can provide enough insight to expose the obfuscation process.

TABLE VI: Gate Error = 0, Link Error = 0 accuracy for Change Prediction Boosted Reconstruction Model across different types of changes.

|       | Level-1          | Level-2          | Level-3        |
|-------|------------------|------------------|----------------|
| c1355 | 81.81 (27 / 33)  | 70.11 (61 / 87)  | 0.00 (0 / 0)   |
| c1908 | 93.54 (29 / 31)  | 57.95 (51 / 88)  | 0.00 (0 / 1)   |
| c2670 | 97.77 (44 / 45)  | 62.12 (41 / 66)  | 44.44 (4 / 9)  |
| c3540 | 90.24 (185 / 205)| 63.67 (163 / 256)| 36.84 (7 / 19) |
| c5315 | 95.95 (308 / 321)| 66.72 (375 / 562)| 46.75 (36 / 77)|
| c6288 | 99.30 (429 / 432)| 79.04 (381 / 482)| 65.21 (30 / 46)|
| c7552 | 89.00 (259 / 291)| 66.66 (394 / 591)| 38.46 (30 / 78)|
| Avg   | 94.3(1281/1358)  | 68.7(1466/2132)  | 46.5(107/230)  |

TABLE VII: Effect of key size variation on accuracy.

|       | Complete Recovery |      |      |      | R-Metric |      |      |      |
|-------|------|------|------|------|------|------|------|------|
| IP    | 0.5x | 1x   | 2x   | 3x   | 0.5x | 1x   | 2x   | 3x   |
| c1355 | 90.0 | 73.3 | 71.2 | 71.6 | 90.0 | 80.7 | 82.3 | 82.3 |
| c1908 | 80.0 | 66.6 | 67.5 | 65.0 | 91.5 | 78.2 | 78.2 | 81.5 |
| c2670 | 90.0 | 74.1 | 77.5 | 80.0 | 96.6 | 79.1 | 86.1 | 86.0 |
| c3540 | 75.0 | 73.9 | 76.5 | 79.7 | 89.0 | 86.8 | 86.6 | 89.4 |
| c5315 | 78.7 | 74.8 | 78.7 | 73.1 | 85.1 | 84.5 | 86.8 | 83.4 |
| c6288 | 91.2 | 87.5 | 83.1 | 86.6 | 96.2 | 94.9 | 92.2 | 93.9 |
| c7552 | 71.8 | 71.1 | 69.8 | 70.3 | 85.8 | 84.6 | 83.9 | 84.6 |
| Avg   | 82.4 | 74.5 | 74.9 | 75.2 | 90.6 | 84.1 | 85.2 | 85.9 |

TABLE VIII: Effect of different key gate insertion heuristics on accuracy.

|       | CS   |      | CY   |      | SLL  |      |
|-------|------|------|------|------|------|------|
| IP    | Acc  | R    | Acc  | R    | Acc  | R    |
| c1355 | 87.5 | 95.7 | 75.0 | 75.0 | 75.0 | 75.0 |
| c1908 | 62.5 | 70.7 | 62.5 | 79.0 | 50.0 | 54.1 |
| c2670 | 75.0 | 83.2 | 75.0 | 75.0 | 87.5 | 87.5 |
| c3540 | 68.7 | 80.0 | 71.8 | 87.3 | 78.1 | 82.2 |
| c5315 | 67.1 | 79.0 | 85.9 | 89.0 | 90.6 | 94.2 |
| c6288 | 79.6 | 89.9 | 85.9 | 93.1 | 75.0 | 86.8 |
| c7552 | 57.8 | 78.9 | 70.3 | 81.6 | 81.2 | 90.5 |
| Avg   | 71.2 | 82.5 | 75.2 | 82.8 | 76.7 | 81.4 |

## VI. IMPLICATIONS AND EFFECTIVENESS

### A. Scalability

*1) Effect of Key Size on Performance:* To observe the effect of variation of number of *Key Size*, we randomly inserted X ($X = [8, 8, 8, 32, 64, 64, 64]$) bit keys in c1355, c1908, c2670, c3540, c5315, c6288, and c7552, respectively. We perform this random insertion on each benchmark separately multiple times to increase our test data size. Next, we generate test data sets with key bits 0.5X ([4,4,4,16,32,32,32]), 2X ([16,16,16,64,128,128,128]) and 3X ([24,24,24,96,192,192,192]) by similar methods. Table VII, shows Gate Error = 0 Link Error = 0 accuracy and R-metric for reconstruction across different *key sizes*. We can see a general increase in reconstruction efficiency as the *key size* decreases. As the ratio $KeyGate/TotalGate$ increases, the chance that key-gates are inserted in close proximity also increases. Increases in this ratio can sometimes lead to more complex transformations of the key-gates after synthesis which makes it harder to recover. For these reasons, we see a drop in reconstruction accuracy as we increase the *key size*. The effect is not persistent as we observe the accuracy is very stable between 1x and 3x. Therefore, increasing the *key size* can slightly improve structural obfuscation, but the design is still very much exposed to a reversion attack.

*2) Effect of Different Key-Gate Insertion Heuristics:* Different heuristics can be used to find the most suitable place to insert the obfuscation gates. In Table VIII, we see the snapshot recovery results for Logic Cone size based insertion (CS), cyclic insertion (CY) and secure logic locking insertion (SLL) [11]. We observe these methods are vulnerable to our snapshot recovery attack and have an R-Metric of over 80%.

*3) Benchmark & Snapshot Size:* Our method does not depend on the size of the benchmark, rather it depends on the size of the snapshots. Therefore, it can easily scale to large designs. We have reported all our results for an output locality size of 3 gates, but our model can be trained and used for predicting a bigger snapshot.

### B. Versatility

The structural information obtained through SAIL can be used in many ways. We discuss some of them next.

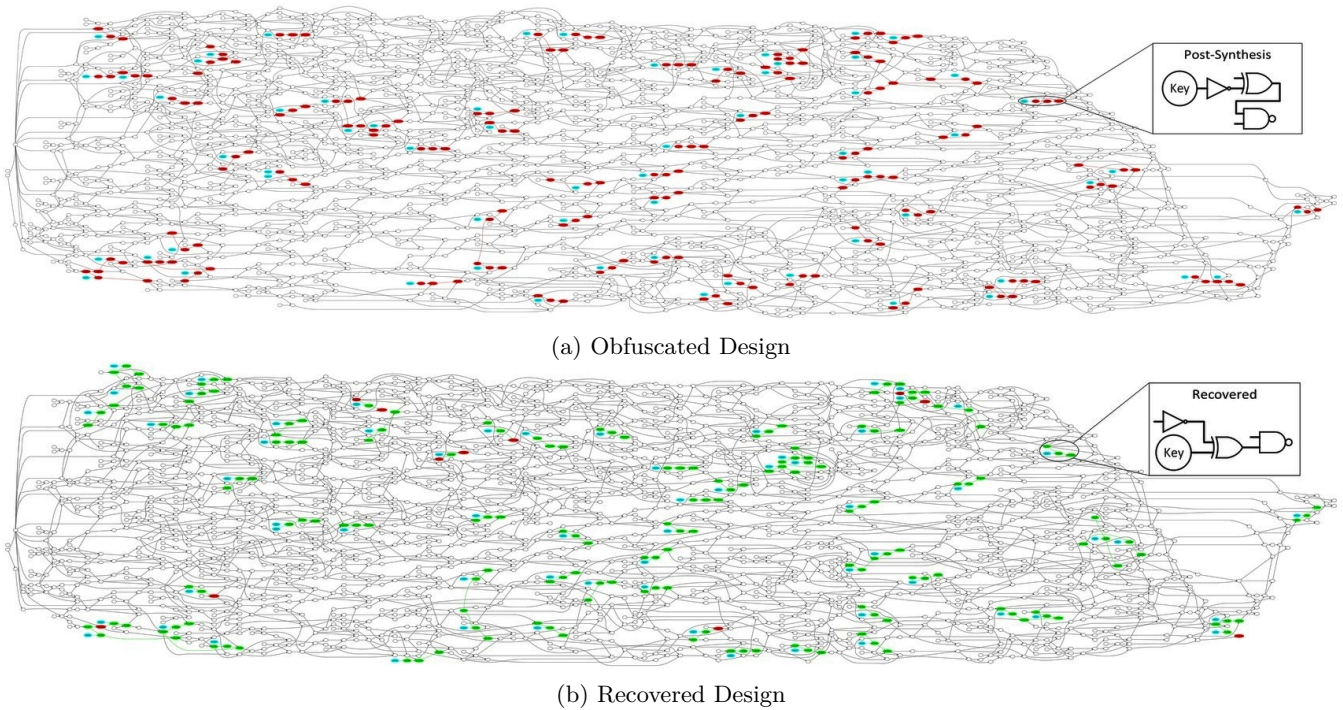

(a) Obfuscated Design

(b) Recovered Design

Fig. 6: Visual representation of SAIL recovery on C6288, with 95% accuracy. Each node is a gate and edges are connections. Green nodes are correctly recovered, and red nodes are incorrectly recovered. Cyan nodes are key inputs.

*1) Key Recovery from Snapshot:* For XOR based locking, recall only a XOR gate is inserted for $keybit = 0$ and a XOR gate followed by an Inverter is inserted for $keybit = 1$. From the locality recovered using SAIL, we can easily predict the value of the key-bit just by observing the gates in the locality.

*2) Aiding Reverse Engineering:* Reverse engineering a netlist to extract its functionality can be severely hampered if a design is obfuscated and re-synthesized. Using SAIL attack, we can recover each key-gate locality in the design and make netlist reverse engineering easier. For example, in Fig. 6, if we can successfully stitch the snapshots, most of the changes that obfuscation introduced can be recovered.

## VII. Conclusion

We have reported a powerful and hitherto unexplored attack modality on logic obfuscation, namely SAIL attack, which exposes a critical vulnerability of these methods. Unlike existing functional attack modes, our attack uses a fundamentally different approach that relies on ML models to retrieve structural changes with high accuracy. The attack becomes possible since these obfuscation methods introduce only small, localized, and predictable transformations in circuit topology. We have presented systematic steps of the attack that lead to the development of an automatic deobfuscation tool. We have studied the efficacy of the attack in terms of both accuracy and computation time for various obfuscation methods and large key sizes. While we report a very high accuracy of structural recovery for a set of benchmarks, we believe enhanced feature selection and training approaches can further improve the effectiveness of the attack. SAIL attack, on one hand, is expected to enable robust security analysis of existing obfuscation methods and on the other hand, help the development of new methods that are resilient to structural attacks. To protect against SAIL attack on obfuscation, we believe, one needs to incorporate the following two features: (1) distributed and global structural changes; and (2) unpredictable patterns of changes. Future work will include exploration of robust SAIL-resistant hardware obfuscation methods.